\documentclass[aps,fleqn,twocolumn]{revtex4}

\newcommand{\bed}{\[}
\newcommand{\eed}{\]}
\newcommand{\beq}{\begin{equation}}
\newcommand{\eeq}{\end{equation}}
\newcommand{\beqa}{\begin{eqnarray}}
\newcommand{\eeqa}{\end{eqnarray}}
\newcommand{\ket} [1] {\vert #1 \rangle}
\newcommand{\bra} [1] {\langle #1 \vert}

\usepackage{graphicx}
\usepackage[mathscr]{eucal}
\usepackage{amsmath}
\usepackage{amssymb}
\usepackage{epstopdf}
\usepackage{epsfig}
\begin{document}

\title{Byzantine Agreement with Two Quantum Key Distribution Setups}
\author{S. Iblisdir and N. Gisin}
\address{GAP-Optique, University of Geneva, 20 rue de l'Ecole-de-M\'edecine, CH-1211, Switzerland }

\date{\today}

\begin{abstract}
It is pointed out that \emph{two separated} quantum channels and three classical authenticated channels are sufficient resources to achieve detectable broadcast.
 \end{abstract}

\maketitle

%Quantum information theory is well-known for having made possible tasks which were believed to be impossible, such as fast factorization \cite{shor95}, and for having unravelled new concepts such as entanglement \cite{nielsen}. Remarkably, quantum information theory has aso provided insight into \emph{classical} information theory, and led to the concepts of bound information \cite{gisi00}, or classical analogs of quantum entanglement \cite{coll01}.

This note is about broadcast (or byzantine agreement), and should be considered as a follow-up of the paper \cite{fitz01}. 

In its simplest form, broadcast is a task involving three parties: a sender, $S$, and two receivers, $R_0$ and $R_1$. The sender holds an input value $x_s \in \mathscr{D}$ ($\mathscr{D}$ denotes some finite domain) and is supposed to send it to the two receivers. The two receivers eventually decide on an input value in $\mathscr{D}$. Amongst $S,R_0$ and $R_1$, one (and at most one) player may be an active adversary and try to stop the two other parties to agree on an input value. The other two parties are said to be honest. 

A protocol achieves \emph{broadcast} if (i) it guarantees that all honest players decide on the same output value $y \in \mathscr{D}$, (ii) $y=x_s$ whenever the sender is honest. One easily sees the difficulties involved in this task if only pairwise (classical) authenticated channels are available. For example, a cheating sender could send different bit values to $R_0$ and $R_1$. Thus $R_0$ and $R_1$ should test the honesty of $S$ during a phase of the protocol where they exchange their input and check whether they match. But if one of the receivers is dishonest, it might happen that, during this verification phase, he sends the other receiver a value different from the input he actually got from the sender. In fact, it is known that, if the only resource available to the three players is pairwise  authenticated channels, broadcast is impossible \cite{lamp82}. However, when quantum channels are available, a variant of the byzantine agreement problem, namely \emph{detectable} broadcast, can be achieved. A protocol is said to achieve detectable broadcast if (i) it achieves broadcast when no player is corrupted. (ii) when one player is corrupted, then either the protocol achieves broadcast, or all honest players abort the protocol. 

Let $\ket{0}, \ket{1}, \ket{2}$ denote an orthonormal basis of a qutrit system (a qutrit is a three-level quantum system). Detectable broadcast can be achieved if the three players share many copies of  the so-called Aharonov state \cite{fitz01}
\bed
\ket{A}=\frac{1}{\sqrt{6}}(\ket{0,1,2}+\ket{1,2,0}+\ket{2,0,1}
\eed
\beq\label{eq:ahar}
-\ket{0,2,1}-\ket{1,0,2}-\ket{2,1,0}),
\eeq
and perform measurements on this state.

Actually, the protocol described in \cite{fitz01} is such that the source of Aharonov states lies at $R_1$'s site. Clearly, Byzantine agreement works as well if  $R_1$ were preparing the state $\ket{A}\bra{A}$ and measuring her qutrit before sending their qutrits to players $R_0$ and $S$. Therefore, instead of using $\ket{A}$, $R_1$ can as well prepare randomly either of the three \emph{two}-qutrit states $\ket{A_0}, \ket{A_1}, \ket{A_2}$, where $\ket{A_0}=2^{-1/2}(\ket{1,2}-\ket{2,1})$ ($\ket{A_1}$ and $\ket{A_2}$ are defined likewise). It is thus clear that Byzantine agreement can be achieved using only \emph{two} quantum channels distributing entangled states.

But even further simplifications can be brought: $\ket{A_0}, \ket{A_1}, \ket{A_2}$ constitute more resources than the three players actually need. An examination of the protocol described in \cite{fitz01} shows that to achieve broadcast, all we need is to meet the five following conditions:
 
\begin{enumerate}

\item $R_0$ and $R_1$ share an n-trit string $K^0 \equiv k_1^0 \ldots k_n^0$; 

\item $S$ and $R_1$ share an n-trit string $K^s \equiv k_1^s \ldots k_n^s$;

\item $\forall j=1 \ldots n$, $k_j^0 \neq k_j^s$;

\item $\forall j=1 \ldots n$, $R_0$ has no information about $k_j^s$ other than $k_j^0 \neq k_j^s$;

\item $\forall j=1 \ldots n$, $S$ has no information about $k_j^0$ other than $k_j^0 \neq k_j^s$.

\end{enumerate}

These conditions can be simply satisfied as follows. $R_1$ uses two quantum channels to distribute private $n$-trit strings (or keys) $K^0$ to $R_0$ and $K^s$ to $S$, by means of a quantum key distribution (QKD) protocol \footnote{Actually, quantum channels can be replaced by classical secret channels, if one allows for computational assumptions.} such as BB84 for example \cite{benn84}. The keys $K^0$ and $K^s$ are supposed to satisfy the above conditions. There are two ways in which $R_1$ could have not done her job properly: (i) the condition 4 or the condition 5 is not satisfied, i.e. 
the keys are not secret, (ii) the condition 3 is not satisfied. 

A violation of condition 4 or condition 5 would imply that $R_1$ cooperates with another player to cheat the third one. This possibility is ruled out by the assumption that there is at most one cheater. A test of condition 3 can be performed by $R_0$ and $S$ (with arbitrarily high statistical confidence), upon $R_0$ sending $S$ a randomly chosen sample of his key.  If this test fails, the protocol should abort.

Thus, either $R_1$ does her QKD job properly (and this fact is acknowledged by $R_0$ and $R_1$) or the protocol will abort. Assuming the first alternative, one can see, repeating all steps of the protocol described in \cite{fitz01}, that detectable broadcast can be achieved, using the three authenticated channels, exactly as if Aharonov states had been used.

\begin{figure}[h]
\begin{center}
\epsfig{figure=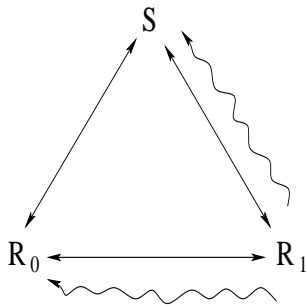,width=40mm,height=40mm}
\caption{Configuration necessary to achieve broadcast. Arrows indicate the direction of the information flow. Wavy lines represent quantum channels, and straight lines represent (classical) authenticated channels.}
\end{center}
\end{figure}

The essential reason why two QKD channels are enough to achieve broadcast is that $R_1$ almost never talks during the protocol described in \cite{fitz01}. Actually, it was already pointed out that no entanglement is necessary to achieve detectable broadcast. A scheme involving three quantum channels and QKD was proposed in \cite{fitz02}. The contribution of this note is to show that only \emph{two} QKD channels are enough.

We thank V. Scarani, R. T. Thew, H. Zbinden, N. Brunner and M. Curty for various discussions. Financial Support from  the Swiss NCCR, and the European projects RamboQ and RESQ are gratefully acknowledged.

%\bibitem{shor95}
%Peter W. Shor, SIAM J.Sci.Statist.Comput.\textbf{26}, 1484 (1997).
Ê%\bibitem{nielsen} M. A. Nielsen and I. L. Chuang, \emph{Quantum Computation and Quantum Information}, Cambridge University Press (2000).
%\bibitem{gisi00} N. Gisin and S. Wolf, LANL report, e-print quant-ph/0005042.
%\bibitem{coll01} D. Collins and S. popescu, LANL report, e-print quant-ph/0107082.

\end{document}